\documentclass[pra,amssymb,twocolumn]{revtex4}

\newcommand{\be}{\begin{equation}}
\newcommand{\ee}{\end{equation}}
\newcommand{\bea}{\begin{eqnarray}}
\newcommand{\eea}{\end{eqnarray}}
\usepackage{epsfig}
\usepackage{graphicx}

\begin{document}

\title{Vortex lattices in dipolar two-component Bose-Einstein condensates}

\author{N. Ghazanfari} \email{nghazanfari@ku.edu.tr} \affiliation{ Department of Physics, Ko\c{c} University, 34450 Istanbul, Turkey} \affiliation{Department of Physics, Istanbul University, 34134 Istanbul, Turkey }
\author{A. Kele\c{s}}\affiliation{Department of Physics, University of Washington, Seattle, Washington 98195-1560, USA}
\author{M.~\"O.~Oktel} \affiliation{Department of Physics,
Bilkent University, 06800 Ankara, Turkey} \affiliation{Laboratory of Atomic and Solid State Physics, Cornell University, Ithaca, New York 14853-2501, USA}

\begin{abstract}
We consider a rapidly rotating two-component Bose-Einstein condensate with short-range s-wave interactions as well as dipolar coupling. We calculate the phase diagram of vortex lattice structures as a function of the intercomponent s-wave interaction and the strength of the dipolar interaction. We find that the long-range interactions cause new vortex lattice structures to be stable and lead to a richer phase diagram. Our results reduce to the previously found lattice structures for short-range interactions and single-component dipolar gases in the corresponding limits.  
\end{abstract}

\maketitle

Cold atom experiments provide the opportunity to study many-particle systems in a highly controlled manner. One of the novel regimes that have gained importance is the study of systems where the particles are interacting significantly through long-range dipolar forces \cite{baranov,fischer,bohn,pu,odell}. The realization of quantum degenerate gas of dipolar bosons and fermions \cite{pfau1,pfau2,lev1,ferlino,lev2} have  given impetus to theoretical study of these systems in various parameter regimes \cite{fischer,bohn,pu,odell}.

The response of Bose-Einstein condensed gases to rotation or an artificial magnetic field has been extensively investigated \cite{ho, mueller1, hui1, oktel,hui2,cooper,moore,mottonen,nitta,kita}. It has been well established that the ground state of a BEC under rotation is a vortex lattice \cite{ho,mueller1,hui1,oktel,hui2,cooper,moore,mottonen,nitta}, and such lattices containing hundreds of vortices have been observed in experiments \cite{ketterle,cornell1}. While the lattice structure for a single-component BEC with short-range interaction is always a triangular lattice \cite{ho}, lattice structures of different symmetry can be obtained either by increasing the number of components in BEC, or by introducing long-range interactions. The phase diagrams of the vortex lattice structures have been calculated for two-component \cite{mueller1} and spin-1 BEC's \cite{kita}. Similarly, the effect of the long-range dipolar interactions on the vortex lattice structure of a single-component BEC have been investigated \cite{hui1,cooper,cooper1}. It is, thus, natural to ask how the long-range interactions modify the phase diagram of the two-component Bose condensates. In this brief report, we calculate the phase diagram of the vortex lattice structures as a function of both the s-wave interactions and the dipolar interactions.  We determine the vortex lattice structures using the method developed in Ref.\cite{mueller1} for two-component condensates and generalized to dipolar interactions in Ref. \cite{hui1}.    

We consider a disk-shaped rapidly rotating two-component Bose-Einstein condensate with contact and dipolar interactions. Each component can be considered as a hyperfine state of the same atom. The orientation of the dipoles are assumed to be fixed by the external field forming the trapping potential. The trap geometry is important in determining the nature of interaction. For the disk-shaped condensates with the dipoles oriented along the symmetry axis, the interaction between atoms is predominantly repulsive. The extent of the cloud along the symmetry axis forms the effective cutoff for the short-range part of the dipolar interaction and can be utilized as a control over the dipolar forces. Similarly, the s-wave interaction strengths can be adjusted by Feshbach resonances, potentially creating a large phase space to explore. As our main aim is to understand the effects of long-range interactions on vortex lattice structure of two-component condensates, we concentrate on a symmetric system where the two components have the same mass, the same density and the same rotation frequency. The dipolar interactions are also assumed to be independent of the component. We calculate the equilibrium vortex lattice structures as a function of the strengths of the short-range intercomponent interaction and the component-independent dipolar interaction.

For a two-dimensional Bose-Einstein condensate confined in an isotropic harmonic trap with a frequency of $\omega$ and rotating at angular frequency $\Omega$ around $z$ axis, the single particle Hamiltonian is 
$ H = \frac{P^2}{2 M}+ \frac{1}{2} M \omega^2 r^2 - \Omega L_z$.
Here $r^2=x^2+y^2$, M is the mass of the particle, and $L_z$ is the total angular momentum in $z$ direction. For such a system $E_{mn}=\hbar(\omega+\Omega)n+\hbar(\omega-\Omega)m+\hbar\omega$, are the energy eigenvalues,  
and the corresponding eigenfunctions are  $\phi_{nm}\propto e^{r^2/2a^2}(\partial_x+i\partial_y)^n(\partial_x-i\partial_y)^m\big(e^{-r^2/a^2}\big)$, where $n\geq 0$ and $m\geq 0$, and $a=\sqrt{\frac{\hbar}{m\omega}}$. As shown in \cite{ho}, when $\Omega$ is large enough, i.e. $\omega-\Omega$ is very small, the system fills the $n=0$ level or the lowest Landau level, known as the mean-field quantum Hall regime. The wavefunction in this regime, for an assembly of cold identical bosons rotating at frequency $\Omega$ can be written as a linear combination of single particle eigenfunctions, $\Psi=f(z) e^{\frac{-r^2}{2a}}$. Here $f(z)$ is a an analytical function of $z=x+iy$. Thus the zeros of $f$ are the positions of the vortices, which will be assumed to 
form an infinite lattice. For a finite condensate, the vortex positions show small deviations from the regular lattice, resulting in a Thomas-Fermi density profile rather than a Gaussian \cite{cornell2}. As we are concerned with the changes in the structure of the lattice, we will neglect these finite-size effects.

For a two-component Bose-Einstein condensate, each component is described with a condensate wave function $\Psi_{i}$, where $i=1,2$. The short-range s-wave interactions and the long-range dipole-dipole interactions are included in the energy functional
\bea \nonumber E[\Psi] &=& \sum_{i=1,2}\int d^2\textbf{r}
\Psi_{i}^* H \Psi_{i} + \sum_{i,j=1,2} \frac{g_{ij}}{2} \int d^2\textbf{r} |\Psi_{i}|^2|\Psi_{j}|^2 \nonumber \\ 
 &+& \sum_{i=1,2} \mu_{i}^2 \int d^2\textbf{r}_{1} d^2\textbf{r}_{2}
|\Psi_{i}(\textbf{r}_{1})|^2 V(\textbf{r}_{1}-\textbf{r}_{2}) |\Psi_{i}(\textbf{r}_{2})|^2  \nonumber\\
& + & \mu_{1} \mu_{2} \int d^2\textbf{r}_{1} d^2\textbf{r}_{2}|\Psi_{1}(\textbf{r}_{1})|^2
V(\textbf{r}_{1}-\textbf{r}_{2})|\Psi_{2}(\textbf{r}_{2})|^2, \eea 
where $g_{ii}=g_{i}=\frac{4\pi\hbar^2 a_{i}}{M}$ and $g_{12}=g_{21}=\frac{4\pi\hbar^2 a_{12}}{M}$ are the the s-wave interaction constants between like and unlike atoms respectively, and $\mu_{i}$'s are the magnitudes of magnetic dipole moment of each component. The magnetic dipole-dipole interaction is 
$V({\bf r}_{1}-{\bf r}_{2})=\frac{\mu_0}{4\pi}\frac{1}{|\bf{r}_{1}-\bf{r}_{2}|^3}$. Here we assume that the magnetic dipoles are parallel to each other, and perpendicular to the line joining the centres of the two dipoles. The densities of both components are considered to be equal. We assume that for the s-wave interactions $g_{1}=g_{2}\neq g_{12}$, and for the dipolar interactions $\mu_{1}=\mu_{2}=\mu$. The wavefunction of each component is normalized such that $\int d^2\textbf{r}|\Psi_{i}|^2=N_{i}$. For a two-component Bose gas in which both components rotate with the same frequency, vortex lattices have the same structure, but one is shifted with respect to another. The wave functions for both components can be introduced by two basis vectors and one relative displacement vector. We assume that $\textbf{B}_{1}$ and $\textbf{B}_{2}$ are the basis vectors of the infinite lattice, and ${\bf r}_{0}= c {\bf B}_1+ d {\bf B}_2$ is the relative displacement of the vortices of different kind. The area of the unit cell is defined to be $v_{c}=|\textbf{B}_{1}\times \textbf{B}_{2}|$. Since the condensate is in the mean field quantum Hall regime, the density $|\Psi({\bf r})|^2$ can be written as a product of a Gaussian and a function $n({\bf r})$ which is periodic under lattice transformation $|\Psi({\bf r})|^2=Ae^{\frac{-r^2}{\sigma^2}} n({\bf r})$ \cite{ho}.
Here $\sigma$ is related to the number of the vortices and is given by 
$\frac{1}{\sigma^2}=\frac{1}{a^2}-\frac{\pi}{v_c}$. 
Periodic function $n({\bf r})$ is expanded as $n({\bf r})=\frac{1}{v_{c}}\sum_{{\bf K}} n_{{\bf K}} e^{i\bf{K\cdot r}}$, where ${\bf K}_i$'s are the reciprocal lattice vectors. 

The presence of $|\Psi|^4$ and $|\Psi_{1}|^2|\Psi_{2}|^2$ in the energy functional leads us to define  $I=\pi \sigma^2 \int d^2\textbf{r}|\Psi_i|^4$ and $I_{12}=\pi \sigma \int d^2\textbf{r} |\Psi_{1}|^2|\Psi_{2}|^2$. In terms of Fourier coefficients, they are given as 
\bea I &=& \sum_{{\bf K},{\bf K}'} \tilde{n}_{{\bf K}}\tilde{n}_{{\bf K}'} e^{\frac{-\sigma^2|{\bf K}+{\bf K}'|^2}{4}}, \\ I_{12} &=& \sum_{{\bf K},{\bf K}'} \tilde{n}_{{\bf K}}\tilde{n}_{{\bf K}'}e^{-i{\bf K}\cdot {\bf r}_{0}} e^{\frac{-\sigma^2|{\bf K}+{\bf K}'|^2}{4}}, \\ \tilde{n}_{{\bf K}} &=& \frac{n_{{\bf K}}}{\sum_{{\bf K}'}n_{{\bf K}'}e^{\frac{-\sigma^2{\bf K}'^2}{4}}}.\eea 
In order to find the optimum vortex lattice structure, we express $n_{\bf K}$'s in terms of the basis vectors. Introducing a complex representation for the basis vectors, $b_{i}=(\hat{\textbf{x}}+i\hat{\textbf{y}})\cdot\textbf{B}_{i}$ and choosing ${\bf B}_{1}$ to lay on the x-axis, the original basis vectors can be written as  ${\bf B}_{1}=b_1\hat{\textbf{x}}$ and ${\bf B}_{2}=b_1(u\hat{\textbf{x}}+v\hat{\textbf{y}})$, where $b_{2}=b_{1}(u+iv)$, and the area of the unit cell becomes $v_{c}=|{\bf B}_{1}\times {\bf B}_{2}|=b_{1}^2 v $. The periodic part of the wavefunction can be chosen as the Jacobi theta function \cite{abramowitz} which has zeros forming a lattice, i.e. $f(z)=\Theta(\zeta,\tau)e^{\pi z^2/2v_c}$, where $\zeta=z/b_1$ and $\tau=b_2/b_1$. Fourier coefficients of $n_{\bf K}$ are easily calculated as
$ n_{{\bf K}}=(-1)^{m_{1}+m_{2}+m_{1}m_{2}} e^{\frac{-v_{c}|{\bf K}|^2}{8\pi}} \sqrt{\frac{v_{c}}{2}}$, and $v_{c}{\bf K}^2=(\frac{2\pi}{v})[(vm_{1})^2+(m_{2}-um_{1})^2]$, 
for ${\bf K}=m_{1}{\bf K}_{1}+m_{2}{\bf K}_{2}$ with ${\bf K}_1$ and ${\bf K}_2$ the  basis vectors of the reciprocal lattice (${\bf K}_1= \frac{2 \pi}{v_c} {\bf B}_2 \times \hat{\bf z}$, ${\bf K}_2= \frac{-2 \pi}{v_c} {\bf B}_1 \times \hat{\bf z}$), and $m_{1}$ and $m_{2}$  integers \cite{mueller1}. For a large number of vortices, the expression for $I$ and $I_{12}$ take simple forms  $ I=\sum_{{\bf K}}|\frac{n_{{\bf K}}}{n_{0}}|^2 $ and  $I_{12}=\sum_{{\bf K}}|\frac{n_{{\bf K}}}{n_{0}}|^2 \cos \textbf{K}\cdot\textbf{r}_{0}$. 
The s-wave interaction energy is
\be E_{s}=\frac{g\rho^2}{\pi\sigma^2}(I+\frac{g_{12}}{g} I_{12}), \ee
with $\rho$ as the average density.

By following the similar steps for dipolar part of the energy expression, 
we write the dipole interaction energy in terms of the relative displacement,  $r=r_{2}-r_{1}$ and the center of mass, $2R=r_{1}+r_{2}$ coordinates, and then integrate with respect to R to obtain
\bea E_{d}=\frac{\rho^2\mu_{0}\mu^2}{4\pi\sigma^2}\sum_{{\bf K}}|\frac{n_{{\bf K}}}{n_{0}}|^2(1+\cos \textbf{K}\cdot\textbf{r}_{0})\left[\frac{1}{\Lambda} -  K\right]. \eea 
Here we define a cutoff, i.e., $\Lambda$, which is related to the thickness of the condensate and regularize the system near to the origin. 
In the limit of a large number of vortices 
the full interaction energy then can be written as 
\be \label{energyint} E_{int}=\frac{\rho^2\mu_{0}\mu^2}{4\pi\sigma^2a}\left[\alpha I+\beta I_{12}-D\right],\ee 
where $D=\sum_{{\bf K}}|\frac{n_{{\bf K}}}{n_{0}}|^2{\bf K}a(1+\cos\textbf{K}\cdot\textbf{r}_{0})$,
$\alpha=\frac{4ga}{\mu_{0}\mu^2}+\frac{a}\Lambda$, and
$\beta=\frac{4g_{12}a}{\mu_{0}\mu^2}+\frac{a}{\Lambda}$.

Since the dipole-dipole interaction is the same for like and unlike atoms, in the energy expression Eq.(\ref{energyint}) $\alpha$ and
$\beta$ can be interpreted as the energy contribution from the intra-component and the intercomponent interaction, respectively. The coefficient $\alpha$ contains the short-range interaction parameter $g$ and it governs the internal behaviour of the individual components. The coefficient $\beta$ contains intercomponent coupling and it determines the lattice offset between the two components. Because of the long-range nature of the dipolar interaction, and its angular dependence a cutoff is needed to regularize the interactions \cite{hui1}. For the pancake harmonic trap considered here, the cutoff parameter $\Lambda$ can be taken as the width of the cloud in the narrow ($\hat{\bf z}$) direction.

\begin{figure}
\begin{center}
\includegraphics[width=8cm]{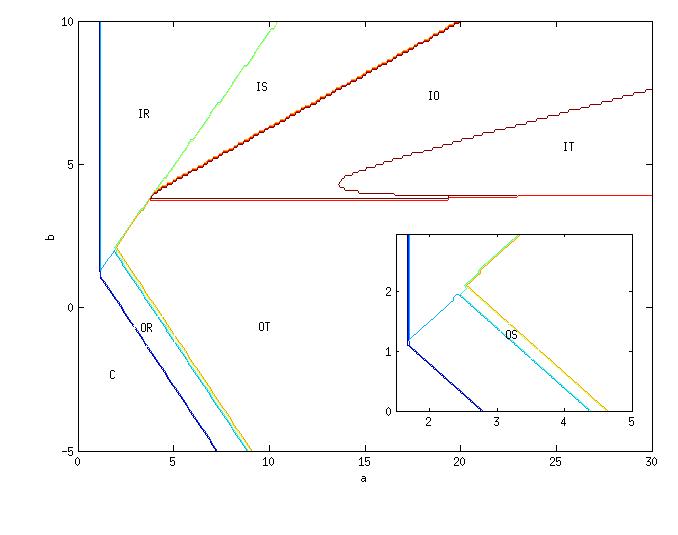}
\end{center}
\caption{(Color online) Phase diagram showing the lattice structures for
different values of the interaction terms $\alpha$ and $\beta$. Here, C corresponds to collapse region, and IR, IS, IO, IT, OR, OT, and OS stand for interlaced rectangular, interlaced square, interlaced oblique, interlaced triangle, overlapped rectangular, overlapped triangle, and overlapped square, respectively. The inset figure indicates the region for overlapped square lattices.}
\label{phasediagram}
\end{figure}

\begin{figure}
\begin{center}
\includegraphics[width=6cm]{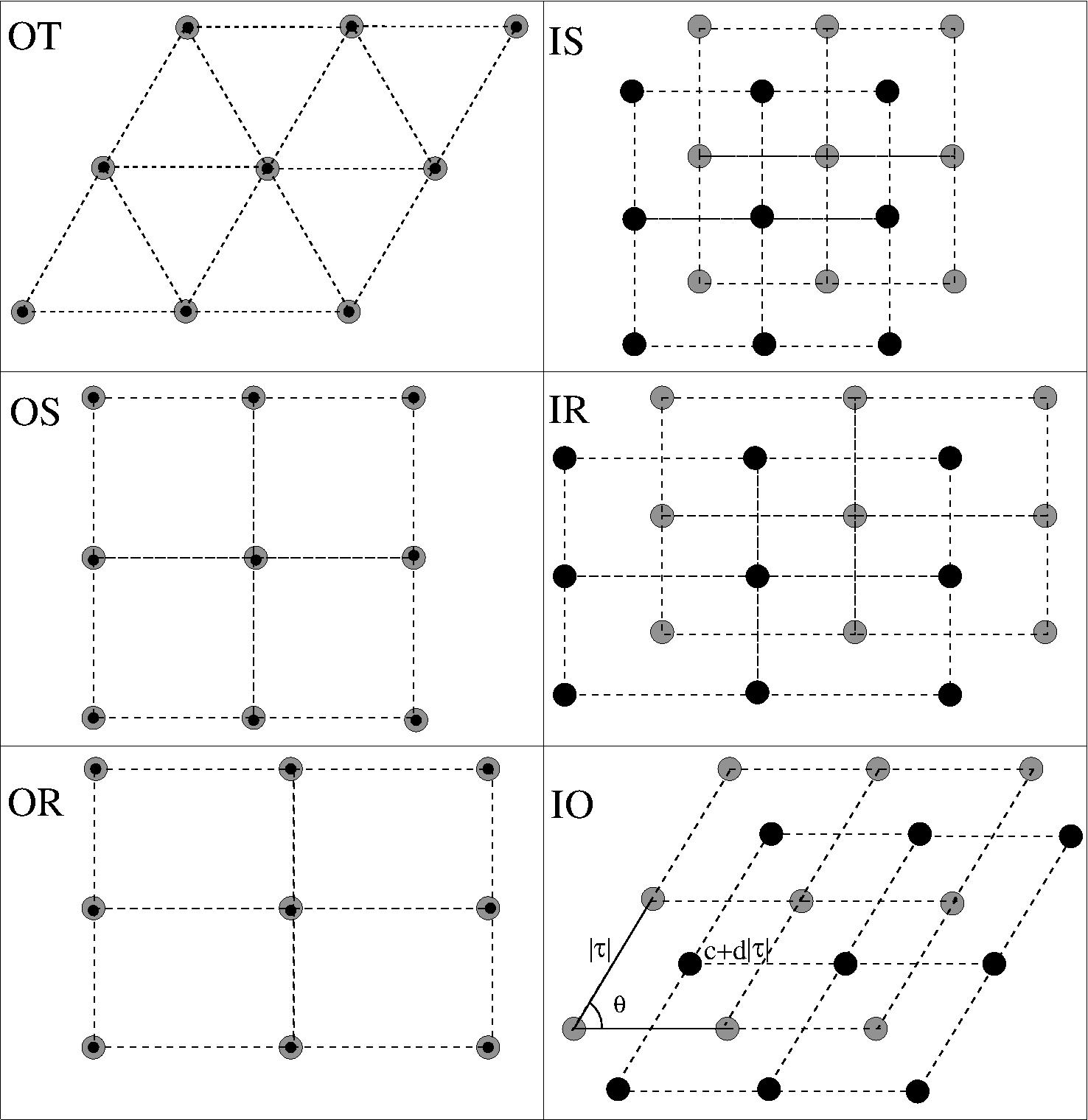}
\end{center}
\caption{Lattice structures for dipolar two-component condensates.
Black and gray dots corresponds to vortices in the two condensates. Lattice structures are defined in terms of aspect ratio $|\tau|$
and lattice angle $\theta$. $c$ and $d$ determine the relative displacement between the two vortex lattices. The calculations are done assuming $c=d$.}
\label{lattice}
\end{figure}

We obtain the phase diagram of the system for different values of $\alpha$ and $\beta$ by minimizing the energy in Eq.(\ref{energyint}) (see Fig.(\ref{phasediagram})). We obtain seven different lattice structures as classified by their symmetry (see Fig.(\ref{lattice})). In three of these phases, the vortices of both components appear at the same points, we call these structures the overlapped lattices. In the remaining four phases the vortices of one component appear at the density maxima of the other component, creating interlaced lattices. As can be expected from our definition of $\beta$, the parameter controlling the intercomponent interaction, these two kinds of lattices are separated roughly by the line $\beta=3$. The dipolar interactions can cause the system to be unstable, which we show as the collapse region in Fig.(\ref{phasediagram}). While all four interlaced lattices have been found for the short-range-interacting systems, the overlapped rectangular and overlapped square lattices are stabilized in a gas with dipolar interactions.

The detailed analysis of the different aspects of the resulting phase
diagram is given in the following points:
i) The attractive interaction causes the condensate to
collapse for $\alpha<1.25$. The condensate collapses even for large
$\beta$ values, since it can not suppress internal fluctuations of each component.
ii) For $\alpha>1.25$, and $\beta<1.25$, the intercomponent
attraction is strong enough to overcome the dipolar repulsion
between unlike atoms, which result in overlapped lattices. By
increasing $\alpha$, the vortex structures undergo a structural
phase transition from overlapped rectangular to overlapped square
lattice and then to overlapped triangle lattices for higher $\alpha$
values.
iii) For $1.25<\alpha<3.70$ and $1.25<\beta<3.70$; the ratio of
these two parameters determines the relative displacement of the two lattices. In this region, when
$\alpha<\beta$, an interlaced rectangular lattice is preferred.  On the
other hand, when $\alpha>\beta$, the minimum energy configuration is an overlapped triangular lattice.
iv) When $\alpha\geq 3.70$ and $\beta \geq 3.70$, only interlaced
lattices exist in the phase diagram, 
since the intercomponent interaction is not attractive in this regime. The
repulsive forces between two different species cause the density minima of
one component to move to the density maxima of the other component. In this region, upon
increasing $\alpha$ the structure of the lattice changes from
interlaced rectangular to interlaced square, oblique, and finally
interlaced triangular.

\begin{figure}
\begin{center}
\includegraphics[width=8cm]{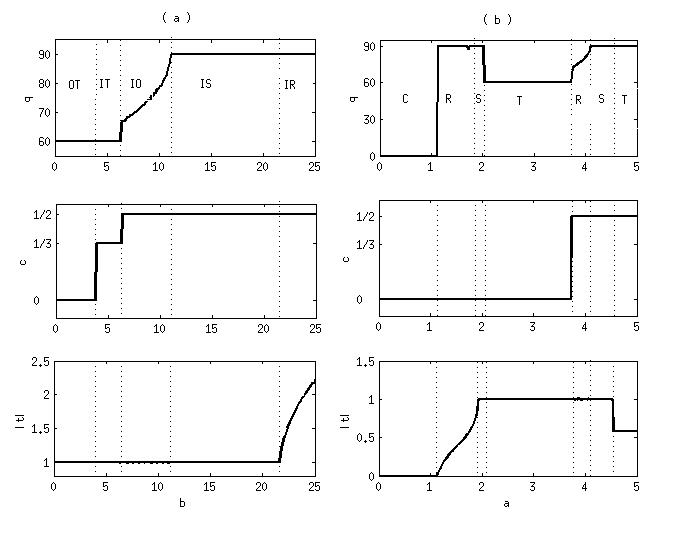}
\end{center}
\caption{The parameters indicating the type of lattice structures as
a function of interaction coefficients $\alpha$ and $\beta$.
(a) The limit for an ordinary two-component condensate, where $\alpha=20$. (b) The limit for a dipolar single-
component condensate, i.e., $\alpha = \beta$. Here $|\tau|$ and $\theta$ are the lattice
parameters  and $c$ indicates the
displacement of the lattice structure of one species with respect
to the other.} \label{mueller-hui}
\end{figure}

Adjusting the strength of the parameters $\alpha$ and $\beta$ also enables us to control the switching between the regime of dominantly
dipolar condensates and the regime of ordinary two-component condensate. One can easily conclude from Eq.(\ref{energyint}) that for small values of $\alpha$ and $\beta$, the dipole-dipole interaction is dominant, and for large values of $\alpha$ and $\beta$, the contact interaction is more prominent.
For large $\alpha$ and $\beta$, the last term $D$ can be ignored
and lattice structures are determined by the ratio $\alpha/\beta$. Thus, the
work reduces to the minimization of the term $J=I+\beta/\alpha I_{12}$.
In the case of dipole-dominant regime, when $g=g_{12}$, two-component gas behaves like a single-component gas. Thus the problem
reduces exactly to the system studied in \cite{hui1} as $\label{hui}E_{int}=\frac{n^2\mu_{0}\mu^2}{2\pi\sigma^2a}\left[\alpha I-D\right]$ . 

To demonstrate the correspondence with works \cite{mueller1} and \cite{hui1}, the phase diagrams along two different lines on the $\alpha\beta$-plane are shown in Fig.(\ref{mueller-hui}). Fig.(\ref{mueller-hui}a) gives the phase diagram for fixed $\alpha$ and changing
$\beta$ that corresponds to two-component condensate with only the short-range interactions as in \cite{mueller1}. Fig.(\ref{mueller-hui}b) shows the phase diagram along and $\alpha=\beta$ line which corresponds to single-component condensate with the short-range and dipolar interactions as in \cite{hui1}.

For $\alpha=\beta$,  we observe the same vortex lattices obtained for the single-component Bose gas with
dipolar interactions (see Fig.(\ref{mueller-hui}b)). However, the correspondence is not straightforward
and requires careful examination: 
a) $\alpha>4.54$: In this region, the triangular vortex lattice is
observed when the two components are considered together. The individual components separately form rectangular lattices but they are interlaced
such that the combined lattice is triangular. It is easy to observe this in Fig.(\ref{lattice}) for IR (interlaced rectangular) lattices.
b) $4.1<\alpha<4.54$: 
The vortices of the two-component condensate form interlaced square lattices, but the combined lattice is again square with a smaller lattice constant.
c) $3.73<\alpha<4.1$:  The two
component gas forms interlaced oblique lattices but the combined
lattice is rectangular, regardless of the angle of the oblique
lattices.
d) $1.12<\alpha<3.73$: The two components form overlapped lattices which both are
identical to the
combined lattice. The combined lattice goes through structural phase transitions in accordance with \cite{hui1}.
e) $\alpha<1.12$: In this region, the condensate collapses.

Compared to the short-range-interacting gas, two new  lattice structures, overlapped square and overlapped
rectangular, are stabilized as a result of dipolar interactions. As the correspondence to the single-component dipolar
gas reveals, these structures are preferred as they maximize the attractive interaction at higher Fourier components of the real 
space density. A two-component gas with short-range interactions can take advantage of the four-fold rotational symmetry only when the two components repel each other,
while the long-range dipolar interaction stabilizes square lattice even for a single-component gas. The phase diagram shows that these two paths of lowering the energy barrier to fourfold symmetry are not mutually exclusive and square vortex lattices can exist for which the two
components are locally attractive. Still, the overlapped square phase is very fragile and covers a relatively small area in the phase diagram. 

In this brief report, we calculated the phase diagram of vortex lattice structures for a two-component BEC in the presence of dipolar interactions. Our results reduce to the ordinary two-component and dipolar single-component vortex lattices in the appropriate limits. Two more lattice structures, the overlapped square and overlapped rectangular lattices, are obtained as a result of dipolar interactions. Experimental observation of these two lattice types would be clear indication of dominant dipolar interactions in a cold atomic gas.
 
\acknowledgments
N.G. is supported by T\"{U}B\.{I}TAK. A.K. is supported by US Department of Energy through the grant  DE-FG02-07ER46452. M. \"{O}. O. is supported by
T\"{U}B\.{I}TAK Grant No. 112T974. M. \"{O}. O. thanks Cornell University for hospitality and 
 the American Physical Society International Travel Award Grant for support. Part of this research was carried out at Aspen Center for Physics, with the support of Simons Foundation and NSF Grant No. 1066293.

\end{document}